\documentstyle[aas2pp4]{article}		

\lefthead{Sakamoto, Baker, Scoville}
\righthead{Gas Dynamics in NGC\,5005}

\newcommand{\etal}{\mbox{\it et al.}}
\newcommand{\Msol}{\mbox{$M_\odot$}}
\newcommand{\Msolpc}{\mbox{$M_\odot$ pc$^{-2}$}}
\newcommand{\Msolyr}{\mbox{$M_\odot$ yr$^{-1}$}}
\newcommand{\kms}{\mbox{km s$^{-1}$}}
\newcommand{\kmskpc}{\mbox{km s$^{-1}$ kpc$^{-1}$}}
\newcommand{\xone}{\mbox{$x_{1}$}}
\newcommand{\xtwo}{\mbox{$x_{2}$}}

\newcommand{\h}{\mbox{$^{\rm h}$}}
\newcommand{\m}{\mbox{$^{\rm m}$}}

\newcommand{\Halpha}{\mbox{H$\alpha$}}
\newcommand{\tm}[1]{\tablenotemark{#1}}
\newcommand{\nd}{\nodata}

\begin{document}

\title{Gas Dynamics in the LINER Galaxy NGC\,5005: \\
	Episodic Fueling of a Nuclear Disk}

\author{
Kazushi Sakamoto\altaffilmark{1,2,3}, 
Andrew J. Baker\altaffilmark{1},
and
Nick Z. Scoville\altaffilmark{1}
}

\altaffiltext{1}{Owens Valley Radio Observatory, Caltech 105-24,
Pasadena, CA 91125}
\altaffiltext{2}{Nobeyama Radio Observatory, Minamisaku, Nagano, 
384-1305, Japan; E-mail(KS): sakamoto@nro.nao.ac.jp}
\altaffiltext{3}{Guest investigator of the RGO Astronomy Data Center}

\vspace{-1cm}

\begin{abstract} 
We report high-resolution CO(1--0) observations in the central 6\,kpc 
(1\arcmin) of the LINER galaxy NGC\,5005 with the Owens Valley Radio 
Observatory millimeter array.  Molecular gas is distributed in three 
components --- a ring at a radius of about 3\,kpc, a strong central 
condensation, and a stream to the northwest of the nucleus but 
inside the 3\,kpc ring.  The ring shows systematic noncircular motions, 
with apparent inward velocities of $\sim 50$ \kms\ on the minor axis.  The 
central condensation is a disk of $\sim 1$ kpc radius with a central 
depression of $\sim 50$ pc radius.  This disk has a molecular gas mass 
of $\sim 2\times10^{9}$ \Msol; it shows a steep velocity gradient and a  
velocity range ($\sim 750$ \kms) 30\% larger than the velocity width 
of the rest of the galaxy.  The stream between the 3\,kpc ring and 
the nuclear disk lies on a straight dust lane seen in the optical.  If 
this material moves in the plane of the galaxy, it lies at a radius of 
$\sim 1$ kpc but has a velocity offset by up to $\sim 150$ \kms\ from 
galactic rotation.
We suggest that an optically inconspicuous stellar bar lying within 
the 3\,kpc ring 
can explain the observed gas dynamics.  This bar is expected to 
connect the nuclear disk and the ring along the position angle of the 
northwest stream.  A position-velocity cut in this direction reveals 
features which match the characteristic motions of gas in a barred potential.
Our model indicates that gas in the northwest stream is on an \xone\ orbit 
at the bar's leading edge; it is falling into the nucleus with a large 
noncircular velocity, and will eventually contribute $\sim 2\times 
10^{8}$ \Msol\ to the nuclear disk.  If most of this material merges with the 
disk on its first passage of pericenter, 
the gas accretion rate during the collision will be 
$\sim 50\,\Msolyr$.  We associate the disk with an inner 
2:1 Lindblad resonance and attribute its large linewidth to favorably 
oriented elliptical orbits rather than (necessarily) to a large central mass.  
The 3\,kpc ring is likely an inner 4:1 Lindblad resonance ring --- or a 
pair of tightly wound spiral arms --- arising at the bar ends.  
Both scenarios can explain the apparent noncircular motions in the ring.
The high rate of bar-driven inflow and the irregular appearance of 
the northwest stream suggest that a major fueling event is in 
progress in NGC 5005.  
Such episodic (rather than continuous) gas supply can regulate  
the triggering of starburst and accretion activity in galactic nuclei.
\end{abstract}

\keywords{ galaxies: dynamics and kinematics ---
	galaxies: ISM ---
	galaxies: spiral ---
	galaxies: active --- 
	galaxies: starburst}

\begin{center}
{\bf To appear in ApJ, 530, \#1 (Feb. 10, 2000). }
\end{center}

\section{Introduction \label{s.intro}}
NGC\,5005 is an Sbc galaxy hosting a low ionization nuclear
emission-line  region (LINER) at its center (Keel 1983).  We list some 
of the system's global properties in Table \ref{t.n5005}.  
Ho \etal\ (1997) propose that the nucleus is accretion-powered, 
on the basis of a broad (1650 \kms\ FWHM) component in a 
fit to the nuclear \Halpha\ line.  Risaliti \etal\ (1999) suggest
on the basis of X-ray observations that the
active nucleus is fully obscured by an absorbing column 
of $N_H > 10^{24}$ cm$^{-2}$.  An alternative driver of 
nuclear emission may be ongoing OB star formation which follows 
a starburst $\sim 10^{9}$ yr ago (Goodrich \& Keel 1986).
On larger scales, NGC\,5005 is classed as SAB (moderately barred)
by de Vaucouleurs \etal\ (1991), although the location of the 
bar-like distortion is not apparent in optical images.  
The galaxy is thought to comprise a binary pair with NGC\,5033, 
a spiral of similar size and redshift at a projected separation 
of $\sim 250$ kpc (for $d = 21.3\,{\rm Mpc}$: Tully 1988).  
Neither galaxy displays distortions to its optical
morphology, suggesting that there has been no significant
recent interaction.  

--- Table \ref{t.n5005} ---

We first observed the molecular gas in NGC\,5005 in the course of a 
survey of 20 nearby spirals (Sakamoto \etal\ 1999a), using 
the Owens Valley Radio Observatory (OVRO) millimeter array.  
We found a strong concentration of molecular gas ($ \sim 10^{9} \Msol $) 
in the central kpc and an [apparently] incomplete ring of gas 
at a radius of 3\,kpc.  The nucleus revealed a very steep velocity 
gradient in CO --- $\Delta V \sim 700$ \kms\ within 10\arcsec\ or 
$\Delta V/\Delta r \gtrsim 800$ \kmskpc\ --- consistent with optical 
observations (Filippenko \& Sargent 1985).  The combination of the  
massive nuclear gas concentration, the large central velocity gradient, 
and the nearly circular ring motivated a more detailed study of this 
galaxy's gas dynamics and their implications for nuclear fueling.
Our original observations had relatively low 
sensitivity and resolution: the $5'' \times 4''$ (0.5 kpc) beam 
only marginally resolved the $\sim 1\,{\rm kpc}$ structure in the 
nucleus.  In this paper, we report high-resolution and high-sensitivity
CO(1--0) observations of NGC\,5005 using data from three 
configurations of the OVRO array.  
These observations include data from the new U configuration,
which with baselines $\sim 400\,{\rm m}$ permits $\sim 1''$ resolution 
at this wavelength.  

\section{Observational Data}

\subsection{OVRO CO(1--0) observations}
Our new aperture synthesis observations of the CO(1--0) emission 
in NGC\,5005 were taken with the OVRO millimeter array between 
February and May 1998, for a single pointing center at 
$\alpha_{1950} =$ 13\h08\m37\fs7 and 
$\delta_{1950} =$ +37\arcdeg19\arcmin29\farcs0.
The array consists of six 10.4\,m telescopes 
equipped with SIS receivers.  Digital correlators were configured to 
cover 336\,MHz (874 \kms) with 4\,MHz (10.4 \kms) resolution. 
Analog correlators recorded 1\,GHz continuum data from the signal 
and image bands.  For gain calibration, we observed the quasar 1308+326 
every half-hour; 3C273 was observed for passband calibration. 
The chopper-wheel method was used to correct for the atmospheric 
transmission. An absolute flux scale (uncertain to $\sim 15\%$) was 
established by comparing the apparent strengths of 1308+326 and Uranus.
We adopt 1.5\,Jy as the flux density for 1308+326 in the spring of 1998, 
an increase over the 1.0\,Jy estimated for the original [survey] 
data from April 1997.  Our observations were obtained in the L, 
H, and [new] U configurations; when combined with the preliminary data, 
we had the equivalent of five full transits with projected baselines from 
10.4\,m to 400\,m.  The visibility data were calibrated with the OVRO MMA 
software (Scoville \etal\ 1993) and mapped with the NRAO AIPS package.

Three combinations of configuration and weighting were used to make maps 
(Table \ref{t.maps}), to which we will refer according to their low 
(4\arcsec), medium (2\farcs7), and high (1\farcs5) resolutions.  
Figure \ref{f.maps} presents all three sets of zeroth and first moment 
maps, while Figure \ref{f.L-chan} shows the low-resolution channel maps.
The fraction of the total flux recovered by each of our maps is estimated
from comparisons of our data with single-dish observations from the 
FCRAO 14\,m and the IRAM 30\,m telescopes 
(Young \etal\ 1995; Braine \etal\ 1993).  
Each data cube was corrected 
for the primary beam response of the OVRO antennas (a Gaussian with HPBW 
= 60\arcsec), convolved with the single-dish beam (45\arcsec\ for 
FCRAO and 23\arcsec\ for IRAM), and sampled at the pointing centers of 
the single-dish observations to obtain values comparable to 
the single-dish fluxes.
Our low and medium-resolution maps recover about 74\% of the
total flux in the central 45\arcsec\ and 93\% 
in the central 23\arcsec.
The high-resolution map recovers 50\% of the total flux in
45\arcsec\ but 75\% in the central 23\arcsec, where
there is a strong concentration of molecular gas.
We made no correction to our data for missing flux.  
We did correct for primary beam attenuation before 
measuring flux from our maps, although this correction 
was not applied to the maps shown in the figures.

--- Table \ref{t.maps} ---

--- Figures \ref{f.maps} -- \ref{f.L-chan} ---

\subsection{Supplementary data}
For comparison purposes, we have obtained data for NGC\,5005 at several other 
wavelengths from archives and the literature.  
Figure \ref{f.INT} shows a Str\"{o}mgren $y$ band ($\lambda = 
550$ nm, $\Delta\lambda = 20$ nm) image obtained with the 2.5\,m Isaac Newton 
Telescope at the Observatorio del Roque de los Muchachos, Spain.  Dust lanes 
in this image trace the spiral arms, which if trailing imply (since the east 
side of the galaxy is receding) that the northwest half of the galaxy is 
nearer to us.  

--- Figure \ref{f.INT} ---

Figure \ref{f.HST-L} shows a WFPC2 image obtained with the Hubble Space 
Telescope (HST).\footnote{
Based on observations made with the NASA/ESA Hubble Space Telescope, 
obtained from the data archive at the Space Telescope Science Institute. 
STScI is operated by the Association of Universities for Research in 
Astronomy, Inc. under NASA contract NAS 5-26555.}
This image was taken in the 
F547M filter ($\lambda = 547$ nm, $\Delta\lambda = 49$ nm). 
Cosmic ray hits were removed and images from the four chips 
of the camera were combined with IRAF and STSDAS. 
The absolute position of the HST image relies on the HST Guide Star Catalog 
and is accurate to about 0\farcs5.  The sharpness of the dust lane  
to the northwest of the nucleus supports the view that this side of the 
galaxy is nearer, and that the spiral arms are trailing.

--- Figure \ref{f.HST-L} ---

NGC\,5005 has been observed with the VLA on several occasions.
A compact central source is only marginally resolved
at $\sim$ 1\arcsec\ resolution.
Table \ref{t.position} lists the measurements of the position 
of the central source.  We also present the mean position of the radio 
source, which coincides well with the dynamical center 
determined from our CO data (\S \ref{s.disk}).

--- Table \ref{t.position} ---

\section{Results}
Our CO maps show an oval ring of $\sim 1'$ diameter
and a strong central condensation, 
confirming our previous observations (\cite{Sakamoto99a}).
In addition to these two structures, we find a ``northwest stream'' between 
the ring and the central peak with 
unusually large noncircular velocities.
We describe each of these three components in order of increasing radius.

\subsection{The nuclear disk \label{s.disk}}
The medium and high-resolution maps of Figure \ref{f.maps} now clearly 
resolve the nuclear gas disk.  
The disk has an extent of $20'' \times 10''$ 
in the medium-resolution map,
elongated along the isophotal major axis of the galaxy (p.a. = 65\arcdeg) as 
determined on large scales.  Its extent is smaller in the 
high-resolution map, probably  
because of lower flux recovery and lower surface-brightness sensitivity.
The emission feature $\sim$ 6\arcsec\ northwest of the nucleus is 
kinematically distinct from the nuclear disk; we discuss it in the next 
section.  The total CO flux of the nuclear disk is
$S_{\rm CO} = 310$ Jy \kms, which corresponds to 
$M_{\rm gas} = 2.3 \times 10^{9}$ for the Galactic conversion factor 
$N_{\rm H_2}/I_{\rm CO} = 3 \times 10^{20}$ cm$^{-2}$ (K \kms)$^{-1}$
(Scoville \etal\ 1987; Solomon \etal\ 1987; \cite{Strong88})
and a factor of 1.36 for heavy elements.
	
We fit the velocity field in the nucleus to a tilted ring model using the 
AIPS task GAL.  The high-resolution data were  
used for the model input; the derived kinematic parameters are listed 
in Table \ref{t.kinem}.  
The inclination and major-axis position angle of the nuclear disk
agree well with those from fits to the isophotes 
of the galaxy as a whole (see Table \ref{t.n5005}).
The slight differences (less than 10\arcdeg\ for both position angle and 
inclination) may be due to noncircular motions in the nuclear disk, 
a warp of the galactic disk, an incorrect assumption for the intrinsic 
axial ratio in the isophote fit,
or statistical errors in both measurements.  
We will adopt the large-scale isophotal values for inclination and position 
angle in the rest of our analysis.
The inferred dynamical center coincides within 0\farcs5 
with the mean position of the nuclear radio source in Table \ref{t.position}.
The CO emission is largely axisymmetric about the
dynamical center, with a slight ($\sim 1''$ or 100 pc) shift of the
emission centroid to the northeast.

--- Table \ref{t.kinem} ---

In the high-resolution map,
CO(1--0) emission weakens at the dynamical center, due to absence of gas or 
higher excitation at the nucleus.
The S-shaped contours in the isovelocity map near systemic
velocity indicate noncircular motions in the nuclear disk.
The noncircular component of the velocity is directed outward on the 
disk's minor axis.  The full range of velocities in the nuclear disk 
(in the medium-resolution data) is 740 \kms,
or 830 \kms\ if corrected for the inclination of 62\arcdeg.
This is about 30\% wider than the HI line width of the galaxy,
$W_{\rm HI} = 556$ \kms\ (not corrected for inclination: \cite{RC3}),
as has been noticed previously (\cite{Braine93}). 
The large linewidth is confined to the nuclear disk, as seen
from a position-velocity cut along its major axis (Figure \ref{f.Hpvs}).
The dynamical mass and the mass fraction of gas within 
radii of 5\arcsec\ (0.5\,kpc) and 30\arcsec\ (3.1\,kpc)
are given in Table \ref{t.mgasmdyn}.
The total mass in the central kpc ($2\times 10^{10}$ \Msol,
derived assuming circular rotation)  
almost certainly overestimates the true dynamical mass, as we will discuss
in \S \ref{s.bar-driven}; but this estimate imposes an upper limit on 
any large mass concentration in the galactic center.
Figure \ref{f.HST-U} shows that the nuclear CO emission coincides with a 
dark arc (on the near side of the galaxy) in the optical.  The center of 
the nuclear disk falls on a conical optical feature whose wide opening 
angle ($\sim 120$\arcdeg) may be 
consistent with outflow out of the galactic plane.

--- Figure \ref{f.Hpvs} ---

--- Table \ref{t.mgasmdyn} ---

--- Figure \ref{f.HST-U} ---

\subsection{The northwest stream}
The emission from the northwest stream between the nuclear disk and the 
3\,kpc ring has anomalously high velocities which are most easily 
seen in channel maps.  In Figure \ref{f.L-chan}, strong emission is seen 
5\arcsec\ -- 10\arcsec\ northwest of the galactic center in at least 
the velocity channels from 938 -- 1062 \kms\ (if not more).
The same structure is seen with better spatial and velocity resolutions 
in the medium-resolution channel maps 
(Figure \ref{f.H-chan}), most clearly from 979 -- 1083 \kms.
The emission is anomalous in that {\em it is redshifted with respect to the
systemic velocity although it appears on the eastern, approaching side of the
galaxy}.  
The (line-of-sight) velocity offset between this stream  
and the galaxy rotation estimated from emission elsewhere 
is as large as $\sim 150$ \kms.  
This feature is definitely real: it is seen in multiple 
subsets of our data at the same location and velocity.  If it traces material 
moving radially in the galactic plane, the implied motion is at 200 \kms\ 
(about half the free-fall velocity) directed {\it inward}.

--- Figure \ref{f.H-chan} ---

The size of the northwest stream in the high-resolution 
map is $5'' \times 3''$ ($0.5 \times 0.3$ kpc), with its major axis running 
in an east-west direction.  Due to missing flux, this map may 
underestimate the 
feature's true extent --- possibly closer to 
$\sim 1$ kpc as seen in the medium-resolution map.
The CO flux (in the medium-resolution map) and the CO-derived mass 
of molecular gas are
$S_{\rm CO} = 20$ Jy \kms\ and $M_{\rm gas} = 1.5 \times 10^{8}$ \Msol\ 
respectively.
The centroid of the component is at $ \sim $ 6 \arcsec\ northwest 
(p.a. = $-70$\arcdeg) of the dynamical center.
The galactocentric radius of this centroid is 1.0\,kpc 
if the stream is moving in the galactic plane.
The velocity gradient in the stream also runs from east to west with higher 
velocities to the east; the total linewidth is roughly 100 \kms.
Figure \ref{f.HST-L} reveals that the northwest stream coincides with a 
dust lane observed optically.  This lane lies to the west of the nucleus and
runs from the nuclear disk in the east to the 3\,kpc ring in the west.
The molecular gas of the northwest stream falls in the middle of the dust lane,
where there is a dark feature in the optical image.

\subsection{The 3\,kpc ring \label{s.ring}}
The low-resolution maps 
(Figures \ref{f.maps} and \ref{f.L-chan})
show a CO ring at a radius of $\sim 30''$ (3 kpc) 
as in our previous observations.
In these maps, the 3\,kpc ring is now found to be nearly  
continuous, completely encircling the center with few gaps.
The ring is nearly circular when deprojected,
as seen from comparing it to the reference ellipse (an appropriately projected 
circle) in the maps.  
Most of the ring emission lies within the annulus from 
20\arcsec\ -- 40\arcsec\ (2 -- 4 kpc), although the outer boundary
is uncertain because of the primary beam attenuation.
The total mass of molecular gas in this annulus is $2.2 \times 10^9$ \Msol.
The mean and peak surface densities in this region are 
$ \langle \Sigma_{\rm gas} \rangle = 40$ \Msolpc\ and
$ \Sigma_{\rm gas}^{\rm peak} = 200$ \Msolpc\ respectively,
after correction for inclination.

The kinematics of the ring seen in the channel maps
(Figure \ref{f.L-chan}) are intriguing.  Emission in each channel is {\it not} 
reflection-symmetric about the major axis of the galaxy (i.e. the major 
axis of the ellipse).  The asymmetry is such that the emission on the 
far/near (i.e. SE/NW) side of the galaxy is consistently shifted towards 
the receding/approaching (i.e. to the NE/SW) side of the 
minor axis. 
This is most obviously seen at systemic velocity in the 938 \kms\ channel.
The emission centroids from the ring and the nucleus are indeed aligned, 
but the alignment differs by 30\arcdeg\ from the galaxy's minor axis.  
Attributing this offset to an error in the position angle determined from the 
galaxy's outer isophotes is unwise: it is hard to make an error as large as 
30\arcdeg\ given the large inclination and undisturbed morphology.\footnote{
So far, no HI observations to determine 
the kinematic major axis have been reported for this galaxy.}
The adopted isophotal position angle is also supported by the fact 
that the highest and lowest-velocity emission from the ring appears 
on the galaxy's 
major axis (see the 658, 689, and 1218 \kms\ channels in Figure 
\ref{f.L-chan}).  The kinematic major axis of the nuclear disk has 
p.a. $\approx$ 70\arcdeg, also in close agreement with the
isophotal major axis.  We instead 
ascribe the asymmetries in the channel maps to systematic noncircular 
motions in the ring.  On the minor axis, these motions appear inward, 
since the emission on the far/near side has lower/higher velocity than 
systemic.  The consistency of the asymmetries from channel to channel 
suggests that the noncircular motion is systematic and exists across the 
ring, although it is not observable near the major axis.

In optical images (Figures \ref{f.INT} and \ref{f.HST-L}), 
the ring appears as a dark arc on the near side of the galaxy.
A dark spiral lane clearly emerges from this part of the ring and winds 
to the southwest at larger radius.  A counterpart to this spiral feature 
on the far side of the galaxy 
would be harder to see, but there are hints of optically bright patches 
emanating from the ring to the northeast.
We do not detect CO emission from the spiral arms located
outside the ring.

\section{Gas dynamics in NGC\,5005}
The contrast in surface density between the nuclear molecular disk 
and the surrounding region within the 3\,kpc ring is striking; it is also 
similar to the gas morphology often seen in the 
nuclei of barred galaxies (Sakamoto \etal\ 1999a).  Beyond its appearance, the 
degree of central concentration in this galaxy is (numerically) one of 
the highest in the Sakamoto \etal\ (1999b) survey, whose results show that 
barred galaxies in general have higher degrees of gas concentration than their 
unbarred counterparts.
We propose that NGC\,5005 contains a small bar within its 3\,kpc ring,
which builds up the central gas condensation and is itself traced by the 
northwest stream.  We locate this bar in a roughly east-west direction so that 
the northwest stream can be produced by a shock along the dust lane at its 
leading edge.  The central depression in the nuclear disk may result 
from concentration
of gas towards $x_2$ orbits at the inner 2:1 Lindblad resonance of the bar, 
although it may also reflect outflow or a variation in excitation.  The 
3\,kpc ring likely arises at the end of the bar, due to 
gas accumulation in either an inner 4:1 Lindblad resonance 
region or a pair of tightly wound spiral arms.  A schematic view of our 
interpretation is shown in Figure \ref{f.illust}.  We present below our 
detailed cases for the existence of the bar\footnote{Eskridge (1999) confirms 
the existence of a bar in near-infrared images of NGC\,5005 obtained for the 
Ohio State University Galaxy Survey.  In addition, 
$K'$ snapshot images from the Palomar 60-inch telescope are consistent 
with a bar of length $\sim 1'$ at a position angle of $\sim$ 80\arcdeg.} 
and the origin of the 3\,kpc ring,
along with a brief discussion of the role played by these structures 
in fueling the nucleus.

--- Figure \ref{f.illust} ---

\subsection{Bar-driven gas dynamics \label{s.bar-driven}}
Figure \ref{f.barPV} shows a position-velocity cut
at a position angle of 80\arcdeg, roughly along the hypothesized bar.
The cut is 5\arcsec\ wide and therefore includes the emission from the 
northwest stream.  The p-v diagram shows two distinct kinematic components.
The component with the steeper velocity gradient arises in the 
nuclear disk; the component with the shallower gradient originates 
in the northwest stream [at ($+5\arcsec, +100$ \kms)]
and parts of the 3\,kpc ring.  
The overall ``tilted X'' pattern is 
characteristic of p-v diagrams in models of bar-driven gas dynamics: 
each kinematic component corresponds to a major family of noncircular 
orbits supported by a barred potential.  Such patterns (also described as 
``figure-of-eight'') have been successfully used to infer the presence of 
bars in the Galaxy (Binney \etal\ 1991) and in edge-on external galaxies 
(e.g. \cite{Garcia95}; \cite{Kuijken95}).  Similar patterns have also 
been observed in a number of barred galaxies with moderate inclinations
(e.g.	M83: Handa \etal\ 1990; 
       NGC 7479: Laine \etal\ 1999; 
       UGC 2855: H\"{u}ttemeister \etal\ 1999).

--- Figure \ref{f.barPV} ---

Figure \ref{f.barmodel} presents a model of gas orbits in a
barred potential which reproduces the main features in the 
position-velocity diagram for NGC\,5005.  
We calculate periodic orbits using the 
damped-orbit model of Wada (1994) and the formulae of Sakamoto \etal\ 
(1999a: Appendix B) for the case in which two inner 2:1 Lindblad 
resonances (ILRs) exist.  We adopt a pattern speed for the bar which 
locates these resonances at radii of 1.1 and 2.1 (in dimensionless units 
for which corotation occurs at 10.0); this choice is motivated by 
considerations which follow in \S \ref{s.oring}.  
The dominant orbits outside the outer ILR and inside the inner ILR are 
parallel to the bar and belong to the \xone\ family.  Between the ILRs, 
orbits which are perpendicular to the bar and belong to the \xtwo\ family 
dominate.  In the left panel of Figure \ref{f.barmodel}, the \xone\ orbits are 
horizontally elongated.  At the ILRs, the position angle of the orbits 
differs by 45\arcdeg\ from the bar major axis because of a damping term in 
the model (Wada 1994; Sanders \& Huntley 1976).
We note that the streamlines in a cloud-based model like this one can have 
sharp turns (giving rise to large velocity widths) at the leading edge of the 
bar (e.g. Lindblad \& Lindblad 1994).  Such features have been attributed by 
other authors (e.g. Regan \etal\ 1999) entirely to the shocks which appear in 
full hydrodynamic simulations.  Our ability to match observations with a 
simpler (cloud-based) approach gives us confidence that its use here is 
reasonable.

--- Figure \ref{f.barmodel} ---

The right panel in Figure \ref{f.barmodel} shows a position-velocity
cut through our model along the bar major axis.
Gas particles on \xtwo\ orbits are plotted as crosses, 
those on \xone\ orbits at the leading edge of the bar are plotted as diamonds,
and those on \xone\ orbits at the trailing edge of the bar are plotted as dots.
The \xone\ gas at the bar's leading edge and the separate \xtwo\ gas
inside the outer ILR (i.e. the diamonds and crosses) form a ``tilted X'' 
pattern very similar to that seen in NGC\,5005.  
Moreover, the gas on the model \xtwo\ orbits has a maximum rotational speed 
much faster than the circular rotation speed for this potential
(shown as a solid line).
This excess arises because the tangential velocity near the pericenter of an 
elliptical orbit is higher than the circular velocity.  Because the major 
axis of the bar is close to the major axis of the galaxy (i.e. the line 
of nodes) in NGC\,5005, material on \xtwo\ orbits will reach pericenter while 
moving nearly along our line of sight.  The bar model thus provides a natural 
explanation for why the linewidth of the nuclear disk exceeds that of the 
galaxy's full H\,I line; we do not need to appeal to a large central mass.

Gas on \xone\ orbits at the leading edge of the bar, producing the 
shallower component of the ``tilted X'' pattern in the model p-v diagram, 
appears to have a velocity {\it less} than circular for NGC\,5005.  This 
deficit arises (as above) because the bar and galaxy major axes are 
nearly (though not exactly) parallel: 
streaming along the bar will occur more in the plane 
of the sky than along our line of sight.  
In the context of this model, the northwest stream seen in NGC\,5005 is clearly
identified as gas moving towards the nuclear disk on an \xone\ orbit at the 
leading edge of the bar. 
More tentatively, we suggest that the emission at ($-20\arcsec, +100$ \kms) 
in Figure \ref{f.barPV} may also trace the leading edge of the bar on an 
\xone\ orbit.  This feature (a possible ``east stream'') can be marginally 
distinguished inside the 3\,kpc ring in the medium-resolution map of Figure 
\ref{f.maps}.
Little emission is seen in the regions of the p-v 
diagram corresponding to \xone\ orbits at the trailing edge of the bar, but 
this is not entirely surprising.  Many barred galaxies show strong 
concentrations of gas at the shocks which form in leading-edge dust lanes.
The likely reason is that gas in a bar is directly channeled to a nuclear
disk through \xone\ orbits at its leading edge.  
After colliding with the nuclear disk near pericenter and entering onto 
\xtwo\ orbits, most of the gas is unavailable to flow 
past the nucleus and trace the trailing edge of the bar.  
This distinction between the 
behaviors of the stellar and gaseous components is emphasized by 
Bureau \& Athanassoula (1999) and Athanassoula \& Bureau (1999) in their 
extensive studies of p-v diagrams for barred galaxies.  Our own model 
confirms that p-v diagrams can sometimes tell us not only whether a 
bar supports two families of orbits, but also whether the gas on the 
\xone\ orbits is at the bar's leading or trailing edge.

\subsection{The origin of the 3\,kpc ring \label{s.oring}}
The 3\,kpc ring has a nearly circular shape but clearly noncircular 
velocities (\S \ref{s.ring}).  Figure \ref{f.Hpvs} shows position-velocity 
cuts along the galaxy's [isophotal] major and minor axes.
The minor-axis p-v diagram clearly shows that the
3\,kpc ring has some degree of noncircular (i.e. radial) motion.  
The gas to the south (north) of the nucleus has negative (positive) 
velocities with respect to the systemic velocity.
The radial motion is directed inward with an amplitude of about 50 \kms\ 
on the minor axis.  We consider two possible scenarios for the origin of 
the ring and its inward motions --- that it constitutes a canonical resonance 
ring at the inner 4:1 Lindblad resonance, and alternately, 
that it is actually a pair of tightly wound spiral arms.
Although we prefer the former hypothesis, the latter can not be
excluded with our current data. 

An inner 4:1 Lindblad resonance region occurs where $\Omega - \kappa/4$ 
(calculated assuming a purely axisymmetric potential) is nearly equal to 
the pattern speed of the bar $\Omega_{\rm p}$ 
(see reviews by Buta \& Combes (1996) and Sellwood \& Wilkinson (1993)); 
it is more widely known as 
the ``ultraharmonic resonance'' (UHR), nomenclature which we will adopt here.
An ``inner'' ring which encircles a bar and falls inside spiral arms
is often identified with the UHR (e.g. Schwarz 1984a; Buta 1986).
Can the UHR ring scenario explain the inward velocities we see in NGC\,5005?
The pattern speed chosen to produce Figure \ref{f.barmodel} places 
the model's UHR at a radius of 6.3; the ratio 
$R_{\rm UHR}/R_{\rm OILR} \sim 3$ successfully matches the ratio
 between the radii of the 3\,kpc ring and the nuclear 
disk.  If the 3\,kpc ring is truly a stable ring rather than a transient 
feature or 
a pair of spiral arms, its constituent 
clouds must follow closed orbits in the frame rotating with the bar.  
In this case, the velocities of material in the ring can be estimated 
from its shape (Buta 1986).
Figure \ref{f.faceon} shows a deprojected version of the CO(1--0) emission 
from NGC\,5005.  
The 3\,kpc ring is roughly circular, but is slightly elongated parallel to the 
bar with an axial ratio $\sim 0.8$, consistent with the average axial 
ratio of 0.82 observed for inner rings in barred galaxies (Schwarz 1984b).  
In a frame rotating with the bar, gas motion will be 
counterclockwise anywhere inside corotation, including at the putative UHR.  
Near the minor axis of the galaxy (i.e. on the vertical line in Figure
\ref{f.faceon}), 
gas on an elongated orbit in the ring will therefore be on its way from 
apocenter to pericenter and will appear to have an inward velocity.

--- Figure \ref{f.faceon} ---

A more quantitative approach is taken in Figure \ref{f.theta-v}, 
in which line-of-sight velocities in the 3\,kpc ring are plotted
with respect to azimuthal angle.
It is evident that circular motion alone fails to reproduce the observations, 
leaving velocity residuals up to $\sim 100$ \kms.
In the UHR model, we can assume that noncircular velocities in 
an elongated ring will be sinusoidal with a phase delay roughly equal 
to the position angle of the bar.
The dashed line in Figure \ref{f.theta-v} shows an elliptical-orbit 
model generated for this assumption which produces reasonable agreement 
with the observations, including the 50 \kms\ noncircular velocity
on the galaxy's minor axis (azimuth $= \pm 90\arcdeg$).
Although we do not attempt parameter fitting for this model, the UHR 
hypothesis is qualitatively consistent with the pattern of noncircular motions
in the 3\,kpc ring.  In this scenario, the spiral arms outside the ring 
need not be tightly wound, and the arm which runs east (west) on large 
scales probably starts from the east (west) end of the bar.

--- Figure \ref{f.theta-v} ---

An alternate explanation for the 3\,kpc ring is that it is not really a ring 
at all, but instead is a pair of tightly wound spiral arms.  Streaming motions
in a spiral density wave will be directed inward along the minor axis for all 
radii inside corotation. 
Numerical simulations (e.g., Roberts 1969, and Roberts \& Hausman 1984)
as well as observations of a number of spiral galaxies
(e.g. M81, Visser 1980; M51, Vogel \etal\ 1993) show 
inward motions on spiral arms near the minor axis 
inside corotation.
The velocity--azimuth angle plot in Figure \ref{f.theta-v} supports one 
element of the tightly wound spiral arm picture.
The residual velocities from circular motion are 
(though somewhat scattered) sinusoidal
with a period of $2\pi$ rather than $2\pi/3$, as expected for any  
ringlike feature inside corotation which is formed by density-wave 
spiral arms 
(Canzian 1993).  A cautionary note for this scenario is the sharply different 
pitch angle displayed by the spiral arms at larger galactocentric radii.  
We would at least expect the eastern (western) spiral arm to start from the 
west (east) end of the bar, if the spiral arms start from the bar's ends
as is usually the case in barred galaxies.  However, this picture may 
still necessitate a zero or (unrealistically) negative pitch 
angle for the ``arms'' comprising the ring (see Figure \ref{f.faceon}).  
Precise determination of the galaxy's inclination
and position angle (e.g. with H\,I mapping) and large field-of-view color 
index maps would both help determine whether the 3\,kpc ring is a resonance 
ring at the UHR or is not a ring at all.

\subsection{Episodic inflow to the nuclear disk}
We have shown that the northwest stream is a 
$\sim 2\times 10^{8}$ \Msol\ gas clump on its way from the 3\,kpc ring 
to the nuclear disk.
At an infall speed of 200 \kms\ and a distance of 1\,kpc from the center, 
most of this mass should soon collide and merge with the nuclear disk; 
part of the stream should already be 
colliding with the outskirts of the disk. 
A crude estimate of the rate of accretion 
to the disk during this collision is ${\dot M} = M/(l/V) \sim 
50\,\Msolyr$, where $l$ is the size of the northwest stream and we assume 
that most of the stream material will merge with the nuclear disk during its 
first passage of pericenter (\S \ref{s.bar-driven}).
The absence of inflowing gas along the full length of 
the bar means that the high accretion rate lasts for only a few Myr,
and that inflow must be largely episodic rather than continuous.
The next episode after the present one may well occur as gas in the 
possible ``east stream'' moves inward.
While a steady supply of tenuous gas may have escaped detection, the 
impending increase of the nuclear disk's mass by $\sim 10$ \% seems likely 
to be more significant.  As this fueling event progresses, we would expect 
it to produce cloud collisions, shocks, and gravitational instabilities in the 
nuclear disk, enhancing the star formation rate and (perhaps) the rate of 
inflow 
to the active nucleus.  If the supply of gas to the nucleus is similarly 
episodic in other galaxies, we have a natural mechanism to regulate the 
intermittent triggering of nuclear starbursts, which Balzano (1983) infers 
have a duty cycle of $\sim 5$ \%.  A connection between kpc-scale fueling and 
nuclear activity is likely to be more indirect at best, since accretion 
luminosity can be significantly boosted by the disruption of an individual 
star (e.g. Eracleous \etal\ 1995) or gas cloud (e.g. Bottema \& Sanders 1986).

\section{Conclusions}
We have observed CO(1--0) emission from the LINER galaxy
NGC\,5005 to study the gas dynamics in the central 6\,kpc (1\arcmin).

\begin{enumerate}
\item 
In its central 6\,kpc, NGC\,5005 has
three distinct molecular gas components --- 
a roughly circular ring of radius $\sim 3$ kpc and gas mass $\sim 2 \times 
10^9$ \Msol;
a central condensation with radius $\sim 1$ kpc and
gas mass $\sim 2\times 10^9$ \Msol;
and a component with size $\sim 0.5$ kpc and gas 
mass $\sim 2\times 10^{8}$ \Msol, located  
about 6\arcsec\ northwest of the nucleus. 
The nuclear disk shows weaker CO(1--0) emission in its 
innermost 100 -- 200 pc.

\item
The 3\,kpc ring shows noncircular velocities, with 50 \kms\
inward motions observed on the minor axis.
The northwest stream has a very large noncircular
motion; its line-of-sight velocity is $\sim 150$ \kms\ different  
from that of the circular galactic rotation.
The nuclear disk has a linewidth $\sim 30$ \% larger than
that of the 3 kpc CO ring and of the entire galaxy's HI emission.
Noncircular motions are also evident in the first moment map 
of the disk.

\item
A stellar bar of length $\sim 5$\,kpc may account for 
these gas distributions and their kinematics.
The bar would run roughly east-west, placing the 
northwest stream and an associated straight dust lane at its leading edge.
The nuclear disk is mainly formed from \xtwo\ orbits elongated
perpendicular to the bar.  The disk's large linewidth is probably due 
to clouds on elliptical orbits which move along our line of sight as 
they reach pericenter, rather than (necessarily) 
to a large mass concentration at the galactic center.
The northwest stream lies on an \xone\ orbit elongated parallel to the bar; 
it is falling towards the nuclear disk, with which it will ultimately collide 
and merge.

\item
The 3\,kpc ring may form at the ultraharmonic resonance of the bar 
or (rather improbably) constitute a pair of spiral arms which originate 
at its ends.  
Noncircular motion would be explained in these scenarios by 
orbital motions in an elongated ring and 
by spiral-arm streaming, respectively.  

\item
The supply of gas to the nuclear disk is episodic rather than
continuous.  
Gas from the outer galaxy will accumulate in the 3\,kpc ring 
and occasionally fall to the nuclear disk, as in the case of the
northwest stream, due to the action of the bar.
The present fueling event will ultimately increase the mass of the 
nuclear disk (by about 10 \%) and impact star formation there within 
several Myr.  
A subsequent fueling event may occur if gas in the possible ``east stream'' 
is truly flowing along the bar.
Such episodic supply of gas may be responsible for triggering 
intermittent nuclear starbursts in other disk galaxies.

\end{enumerate}

A future paper (Baker \etal\ 1999) will present further analysis
of the nuclear disk and its relation to fueling of the LINER, based on 
1\arcsec\ resolution maps of multiple molecular lines.

\acknowledgements
We are grateful to the OVRO engineering and technical staff for their help 
with the observations, especially leading up to and during the rollout of the 
new U configuration.   We thank the referee, William Keel, for very useful 
comments and suggestions.  We also acknowledge helpful discussions with Tamara 
Helfer, Colleen McDonald, Alice Quillen, Michael Regan, and Wal Sargent.
Paul Eskridge graciously shared results from the OSU Galaxy Survey 
prior to publication; Amaya Moro-Martin, Alberto Noriega-Crespo, and Leonardo 
Testi generously undertook the $K'$ imaging at Palomar.
The OVRO millimeter array is funded by NSF grant AST 
96-13717 and the K. T. \& E. L. Norris Foundation.  
K.S. was supported in part by a JSPS grant-in-aid.  A.J.B. was supported in 
part by an NSF Graduate Research Fellowship.  
This research has made use of the NASA/IPAC Extragalactic Database
(NED) which is operated by the Jet Propulsion Laboratory, California
Institute of Technology, under contract with the National Aeronautics
and Space Administration.


\clearpage

\figcaption{
CO(1--0) zeroth and first moment maps of NGC\,5005. 
Synthesized beams are shown in the lower left corners; large crosses mark the 
dynamical center at $\alpha_{1950} =$ 13\h08\m37\fs67 and 
$\delta_{1950} =$ +37\arcdeg19\arcmin28\farcs34.  (Top row) Low-resolution 
(4\arcsec) maps.  The ellipse is centered on the dynamical center, and 
has major and minor axes of 60\arcsec\ and $60''\cos62\arcdeg$
with a position angle of 65\arcdeg, which corresponds to a circle in the 
galactic plane after deprojection.  
(Middle row) Medium-resolution (2\farcs7) maps.
(Bottom row) High-resolution (1\farcs5) maps of the nuclear gas disk.
The small crosses mark the mean position of the radio continuum nucleus.
\label{f.maps}}

\figcaption{ 
Low-resolution channel maps of CO(1--0) emission from NGC\,5005.
Each panel is labeled with the LSR velocity in \kms.
The synthesized beam is shown in the lower left corner of the first panel.
The ellipse is identical to that shown in Figure \ref{f.maps}.
\label{f.L-chan}}

\figcaption{
Optical image of NGC\,5005 ($\lambda = 550$ nm). 
North is to the top and east is to the left.
\label{f.INT}}

\figcaption{
The low-resolution CO(1--0) map overlaid on the HST $V$-band image of
NGC\,5005.  Dark dust lanes and molecular gas correlate well.
\label{f.HST-L}}

\figcaption{
Position--velocity cuts along the major and minor axes of the galaxy 
(p.a. 65\arcdeg\ and 155\arcdeg, respectively), made from
the medium-resolution data.
Position is measured from the dynamical center, 
and velocity is relative to 932.8 \kms.
\label{f.Hpvs}}

\figcaption{
The central $14'' \times 14''$ of NGC\,5005.
The high-resolution CO(1--0) map is overlaid on the HST $V$-band image.
The cross marks the radio nucleus.
\label{f.HST-U}}

\figcaption{
Medium-resolution CO(1--0) channel maps for channels containing emission 
from the northwest stream, the centroid of which is shown as a cross.
Each panel is labeled with the LSR velocity in \kms.
The synthesized beam is shown in the lower left corner of the first panel.
The ellipse is identical to that shown in Figure \ref{f.maps}.
\label{f.H-chan}}

\figcaption{ 
Schematic view of the gas distributions in NGC\,5005.
The 3\,kpc ring, the nuclear molecular disk, and the northwest stream 
are seen in CO, as is a possible ``east stream.''
A dust lane between the nuclear disk and the 3\,kpc ring (coincident with the 
northwest stream) and 
the outer spiral arms are suggested from dark lanes
in optical images.
A bar or oval distortion shown as a dashed ellipse is
inferred from the gas distribution and kinematics.
The major axis of the galaxy has p.a. 65\arcdeg; the northwest
side of the galaxy is near to us; and the sense of rotation is 
counterclockwise.
\label{f.illust}}

\figcaption{
Position--velocity cut along the bar (p.a. 80\arcdeg) through the 
dynamical center, made from the low-resolution data.  The cut is 
5\arcsec\ wide and contains the northwest stream
at about ($+5''$, $+100$ \kms) and the possible ``east stream''
at about ($-20''$, $+100$ \kms).
The velocity zero-point is 937.9 \kms.
\label{f.barPV}}

\figcaption{
(Left)
Model orbits in a barred potential, in a frame rotating with the bar.
The bar runs horizontally (along the dashed line) and rotates counterclockwise.
The potential is 
$ \Phi(r, \theta) = 
  - \left( 1 + \varepsilon \cos 2\theta \right)
  \log(1 + r^2) / 2 $,
where $(r, \theta)$ are polar coordinates rotating with
the bar and the strength of the bar is $\varepsilon = 0.04$.
The pattern speed of the bar is 0.1, which places 
the inner and outer ILR, UHR, and corotation radii at 1.1, 2.1,
6.3, and 10, respectively.
The orbits are calculated using a damped orbit model (Wada 1994).
Horizontally elongated orbits are \xone\ orbits, 
while vertically elongated ones are \xtwo\ orbits.
(Right)
Position--velocity cut (of full spatial width 8) through the model galaxy 
at left, along the bar major axis.  
The solid line shows the circular velocity curve.
Gas particles on different orbits are plotted with different symbols: 
crosses are on \xtwo\ orbits, diamonds are on \xone\ orbits at the leading 
edge of the bar, and dots are on \xone\ orbits at the trailing edge of the bar.
Gas on \xtwo\ orbits and on \xone\ orbits at the leading edge of the bar (i.e. 
crosses and diamonds) form a ``tilted X'' pattern like that seen in Figure 
\ref{f.barPV} for NGC\,5005.  Velocities have been corrected for inclination, 
assuming that the line of nodes and the bar major axis differ by 30\arcdeg\ 
in the plane of the galaxy.
\label{f.barmodel}}

\figcaption{
Face-on view of NGC\,5005, made from the low-resolution map. 
Deprojection assumed $i = 62\arcdeg$ and P.A. = 65\arcdeg.
Primary-beam correction has been applied.
In this figure, the line of nodes is horizontal, 
the reference circle has a 30\arcsec\ radius,
and the line at p.a. $\approx -45\arcdeg$ shows the approximate
position angle of the bar (east-west in the sky plane).
The approaching side of the galaxy is to the right, 
and the near side is at top. 
Gas rotates counterclockwise.
\label{f.faceon}}

\figcaption{
Velocities on the 3\,kpc ring are plotted as a function of azimuthal angle
$\theta$ in the galactic plane, where 
$\theta$ is measured counterclockwise 
from the receding major axis (p.a. = 65\arcdeg).
Data points (asterisks) are mean line-of-sight velocities for the
ring ($R = 25''$ -- $35''$) with $\pm 1 \sigma$ standard deviations,
measured from the low-resolution CO data.
The solid curve shows velocities expected for circular rotation
with azimuthal velocity $V_{\rm az} = 295$ \kms\ in the galactic plane
($V_{\rm sys} = 952$ \kms).
The dotted curve shows the expectation for an oval orbit with
$V_{\rm rad} = -60 \sin2(\theta - 30\arcdeg)$ and
$V_{\rm az} = 295 - 30 \cos2(\theta - 30\arcdeg)$.
At bottom are the residual velocities for each model.  The 30\arcdeg\ phase 
offset corresponds to the difference between line of nodes and bar major axis 
which was used to correct Figure \ref{f.barmodel} for inclination.
\label{f.theta-v}}


\clearpage

\begin{deluxetable}{lcr}
\tablecaption{Parameters of NGC\,5005 \label{t.n5005}}
\tablewidth{0pt}
\tablehead{
 \colhead{Parameter} & 
 \colhead{Value } &
 \colhead{Ref.} 
}
\startdata
Hubble type 				& SAB(rs)bc	& (1) \nl
Distance [Mpc]				& 21.3		& (2) \nl
P.A. (isophotal)	[\arcdeg]	& 65		& (1) \nl
inclination (isophotal) [\arcdeg]	& 62		& (1) \nl
$V_{\rm sys}$		[\kms]\tm{a}	& 952		& (1) \nl
$B_{T}^{0}$		[mag]		& 10.19		& (1) \nl
$S_{\rm CO(1-0)}(45'')$ [Jy \kms]	& $697 \pm 90$	& (3) \nl
$S_{\rm CO(1-0)}(23'')$ [Jy \kms]	& $435 \pm 31$	& (4)
\enddata
\tablecomments{ References are 
(1) de Vaucouleurs \etal\ (1991);
(2) Tully \etal\ (1988);
(3) Young \etal\ (1995), 45\arcsec\ (FHWM) beam centered at
    $\alpha_{1950} =$ 13\h08\m37\fs6 and
    $\delta_{1950} =$ +37\arcdeg19\arcmin25\arcsec;
(4) Braine \etal\ (1993), 23\arcsec\ (FWHM) beam centered at 
    $\alpha_{1950} =$ 13\h08\m37\fs7 and
    $\delta_{1950} =$ +37\arcdeg19\arcmin28\farcs5.
}
\tablenotetext{a}{HI velocity, converted to the LSR velocity 
in the radio convention.}
\end{deluxetable}


\begin{deluxetable}{ccccrccccc}
	\footnotesize
\tablecaption{Parameters of CO(1--0) maps \label{t.maps}}
\tablewidth{0pt}
\tablehead{
 \colhead{Resolution} & 
 \colhead{Configs \& Weighting} &
 \multicolumn{3}{c}{Beam} &
 \colhead{$V_{\rm res}$} &
 \multicolumn{2}{c}{$\sigma$} &
 \colhead{$f_{45''}$} &
 \colhead{$f_{23''}$} \nl
 \colhead{(1)} &
 \colhead{(2)} &
 \multicolumn{3}{c}{(3)} &
 \colhead{(4)} &
 \multicolumn{2}{c}{(5)} &
 \colhead{(6)} &
 \colhead{(7)} \nl
 \colhead{ } &
 \colhead{ } &
 \colhead{\arcsec} &
 \colhead{\arcsec} &
 \colhead{\arcdeg} &
 \colhead{\kms} &
 \colhead{mJy bm$^{-1}$} &
 \colhead{mK} &
 \colhead{\%} &
 \colhead{\%} 
}
\startdata
Low	& L+H; NA 	         & 4.57 & 3.46	& $-21.25$	& 31.1 	&  9.9	& 58	& 75	& 92	\nl
Medium	& L+H+U; UN, robust 2.0  & 2.88	& 2.44	& $-18.88$	& 10.4	& 13.8	& 181	& 73	& 94	\nl
High	& L+H+U; UN, robust -0.5 & 1.55	& 1.42	& $-89.95$	& 20.7	& 16.8	& 705	& 47	& 75	\nl
\enddata
\tablecomments{
Col. (2): L,H,U are the observing configurations. UN = uniform weighting, NA = natural weighting.
Col. (3): Full width at half maximum along the major and minor axes of the beam, and
position angle of the major axis.
Col. (4): Velocity width of each channel map.
Col. (5): $1\sigma$ noise level expressed in intensity and brightness temperature.
Col. (6): Fraction of flux recovered in the data cube, out of total seen by 
the FCRAO 14\,m telescope in a 45\arcsec\ FWHM beam (Young \etal\ 1995).
Col. (7): Fraction of flux recovered in the data cube, out of total seen by 
the IRAM 30\,m telescope in a 23\arcsec\ FWHM beam (Braine \etal\ 1993).
}
\end{deluxetable}

\clearpage

\begin{deluxetable}{lcclcclccl}
\tablecaption{Position of the nucleus \label{t.position}}
\tablewidth{0pt}
\tablehead{
 \colhead{\#} & 
 \multicolumn{3}{c}{R.A. (B1950)} &
 \multicolumn{3}{c}{Dec. (B1950)} &
 \colhead{Resolution} &
 \colhead{$\lambda_{\rm obs}$} &
 \colhead{Ref.} \nl
 \colhead{ } &
 \colhead{$^{\rm h}$} & 
 \colhead{$^{\rm m}$} &
 \colhead{$^{\rm s}$} &
 \colhead{\arcdeg} & 
 \colhead{\arcmin} &
 \colhead{\arcsec} &
 \colhead{\arcsec} &
 \colhead{cm} &
 \colhead{ }  
}
\startdata
1	& 13 & 08 & 37.68	& +37 & 19 & 28.77	& 1	& 6	& (a) \nl
2	& 13 & 08 & 37.67	& +37 & 19 & 28.4	& 1.5	& 20	& (b) \nl
3	& 13 & 08 & 37.69	& +37 & 19 & 28.6	& 1.0	& 6,20	& (c) \nl
4	& 13 & 08 & 37.651	& +37 & 19 & 28.80	& 5	& 20	& (d) \nl
5	& 13 & 08 & 37.673	& +37 & 19 & 28.64	& \nd	& \nd	& (e) \nl
6	& 13 & 08 & 37.67	& +37 & 19 & 28.34	& 1.5	& 0.3	& (f) 
\enddata
\vspace{-8mm}
\tablecomments{Sources are 
(a) van der Hulst \etal\ (1981);
(b) Condon \etal\ (1990);
(c) mean of 6 and 20\,cm positions in Vila \etal\ (1990);
(d) VLA/FIRST survey (White \etal\ 1997), converted from J2000 to B1950;
(e) arithmetic mean of 1,2,3, and 4;
(f) this paper: dynamical center determined from the high-resolution CO(1--0) data.}
\end{deluxetable}


\begin{deluxetable}{rc}
\tablecaption{Kinematic parameters of the nuclear disk \label{t.kinem}}
\tablewidth{0pt}
\tablehead{
 \colhead{Parameter} & 
 \colhead{Value }
}
\startdata
Dynamical center $\alpha_{1950}$ & $ 13\h08\m37\fs67 $ \nl
		$\delta_{1950}$ & $ +37\arcdeg19\arcmin28\farcs34 $ \nl
P.A. 		[\arcdeg]	& 70	 \nl
inclination     [\arcdeg]	& 53	 \nl
$V_{sys}$	[\kms]		& 928\tablenotemark{a}	\nl
Velocity range	[\kms]		& 565 -- 1301\tablenotemark{b} \nl
$\Delta V$	[\kms]		& 736\tablenotemark{b} 
\enddata
\vspace{-8mm}
\tablecomments{All values except the velocity range and the linewidth
are determined from a fit to the high-resolution first moment map in Figure 
\ref{f.maps}.  The northwest stream is excluded from the fit.
For the remainder of our analysis, 
we adopt the very similar isophotal position angle and inclination 
in Table \ref{t.n5005}.}
\tablenotetext{a}{LSR velocity defined in the radio convention.}
\tablenotetext{b}{Includes channels with emission at the $3 \sigma$ level 
in the medium-resolution data cube.}
\end{deluxetable}


\begin{deluxetable}{rcccc}
\tablecaption{Gas and dynamical masses \label{t.mgasmdyn}}
\tablewidth{0pt}
\tablehead{
 \multicolumn{2}{c}{Radius\tm{a}} & 
 \colhead{$M_{\rm dyn}$\tm{b}} &
 \colhead{$M_{\rm gas}$\tm{c}} &
 \colhead{$M_{\rm gas}/M_{\rm dyn}$} \nl
 \colhead{\arcsec} &
 \colhead{kpc} &
 \colhead{$10^{10}$ \Msol} &
 \colhead{$10^{9}$ \Msol} &
 \colhead{\%}
}
\startdata
 5 &	0.5 &	2.1 &	1.4 &   7 \nl
30 &	3.1 &	6.3 &	4.0 &	6 
\enddata
\tablenotetext{a}{Radius in the galactic plane.}
\tablenotetext{b}{Rotation velocities are measured from the 
major axis position-velocity diagram in Figure \ref{f.Hpvs}. 
The dynamical mass in the central 5\arcsec\ is probably overestimated 
because of noncircular motion (see \S \ref{s.bar-driven}).}
\tablenotetext{c}{CO fluxes are measured from the medium-resolution data
for $R \leq 5\arcsec$ and from the low-resolution data for $R \leq 30\arcsec$.}
\end{deluxetable}


\setcounter{figure}{0}

\begin{figure*}[t]
{\hfill\epsfxsize=17cm\epsfbox{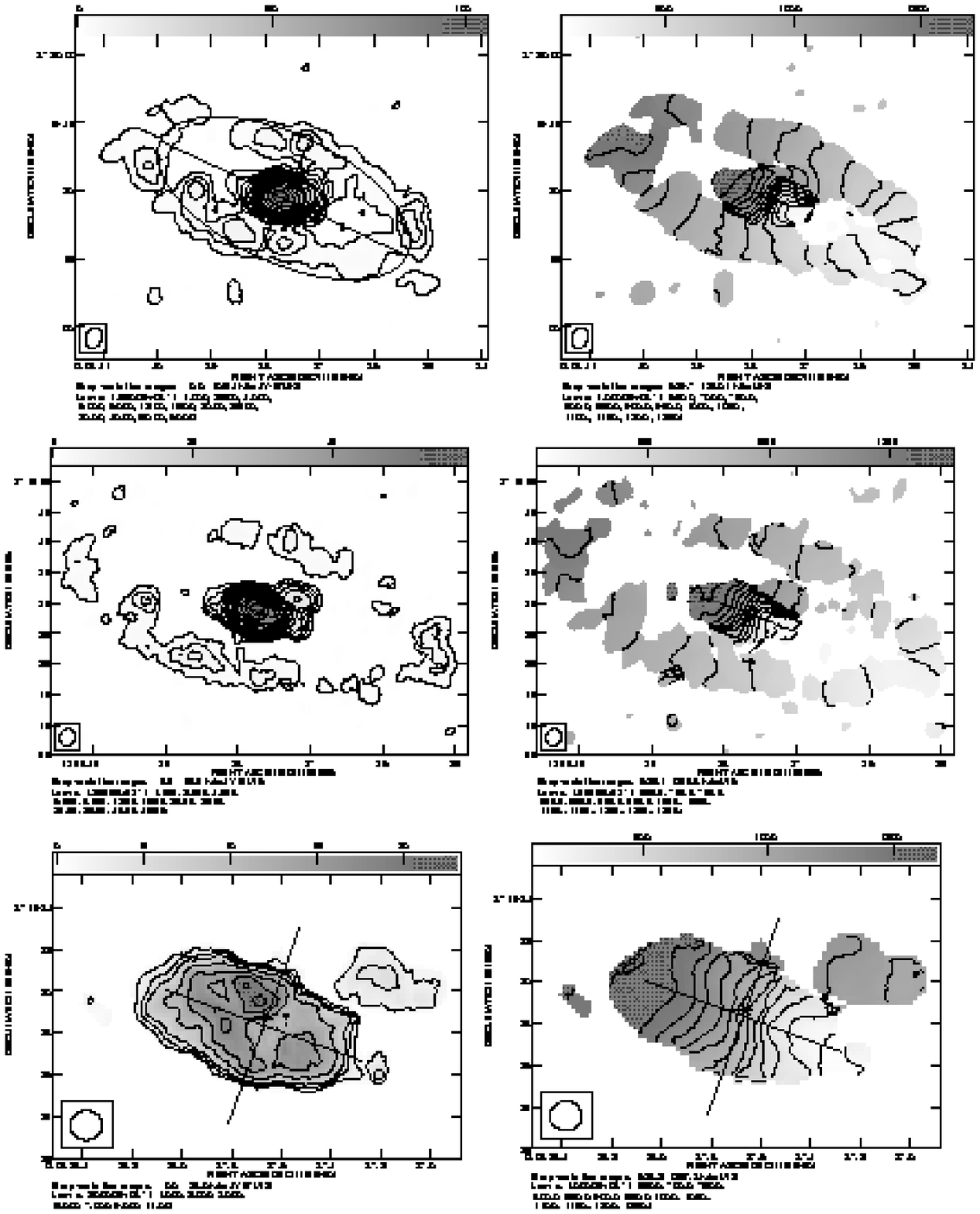}\hfill} 
\vspace{1cm}
\caption{Quality of this figure was lowered for astro-ph.
A PS preprint with the full-resolution figures is avaiable from
{\tt http://www.ovro.caltech.edu/mm/science/science.html} . }
\end{figure*}

\clearpage

\begin{figure*}[t]
{\hfill\epsfxsize=17cm\epsfbox{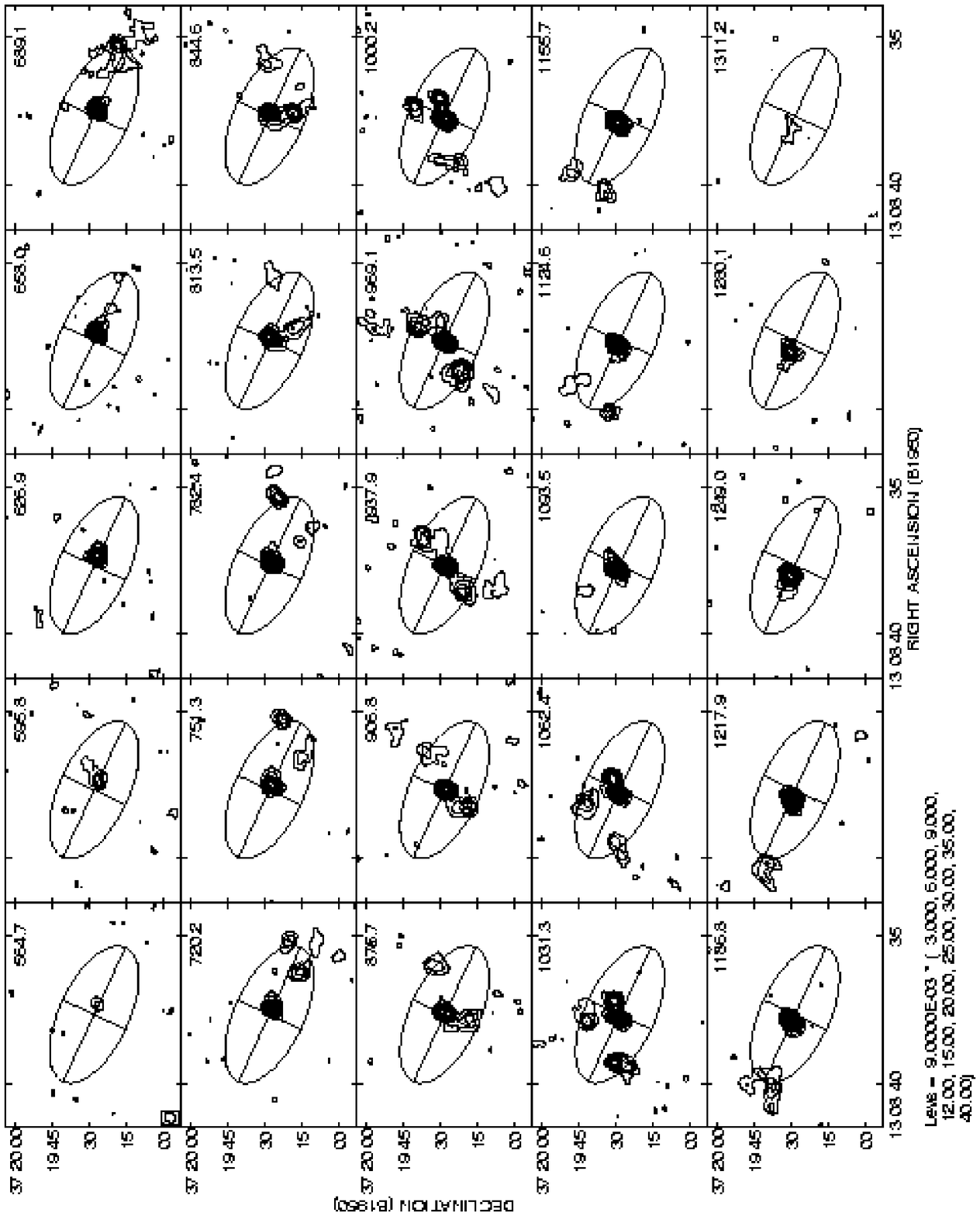}\hfill}
\vspace{1cm}
\caption{Quality of this figure was lowered for astro-ph.
A PS preprint with the full-resolution figures is avaiable from
{\tt http://www.ovro.caltech.edu/mm/science/science.html} .  }
\end{figure*}

\clearpage

\begin{figure*}[t]
{\hfill\epsfxsize=16cm\epsfbox{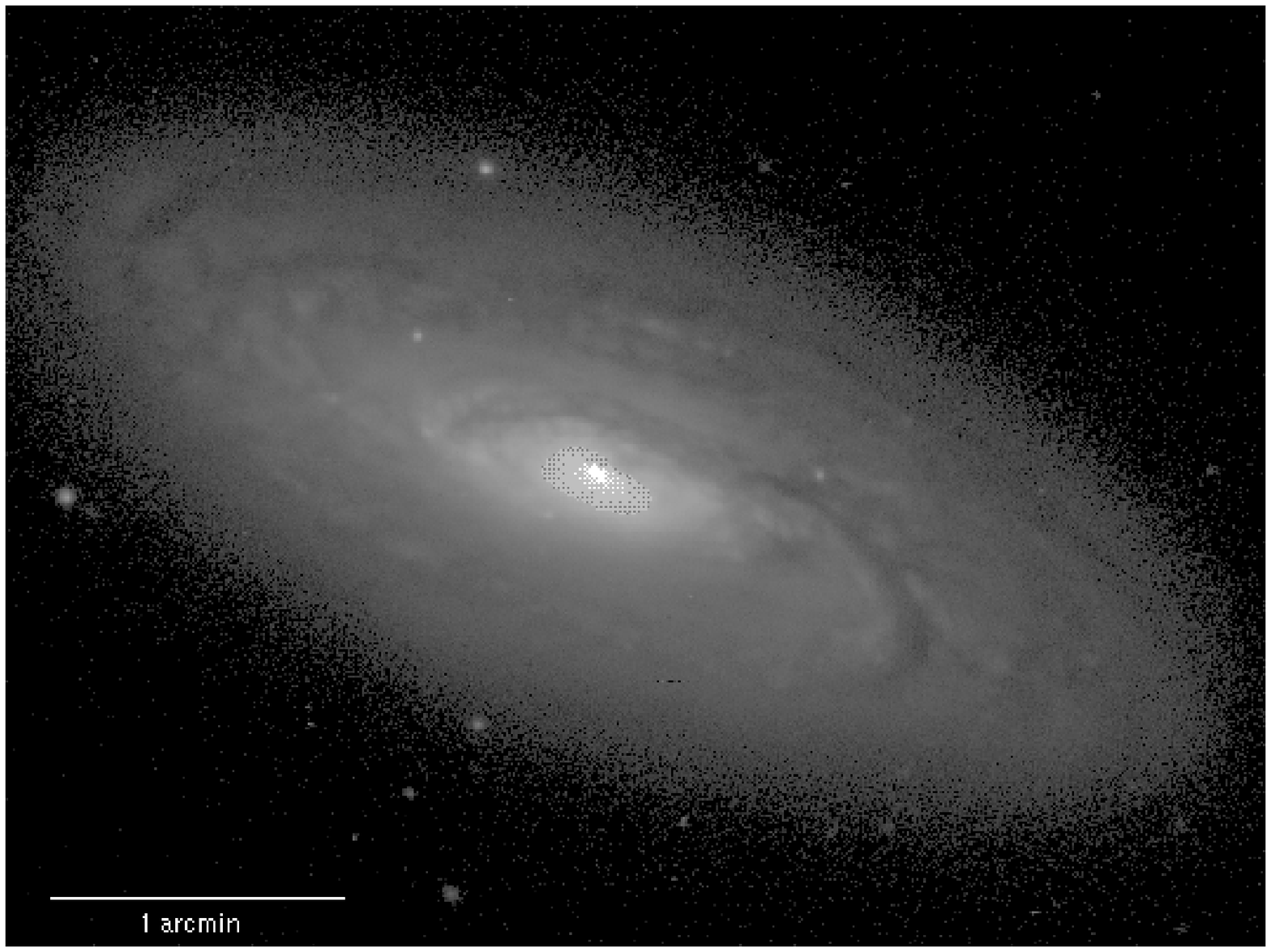}\hfill}
\vspace{1cm}
\caption{Quality of this figure was lowered for astro-ph.
A PS preprint with the full-resolution figures is avaiable from
{\tt http://www.ovro.caltech.edu/mm/science/science.html} . }
\end{figure*}

\clearpage

\begin{figure*}[t]
{\hfill\epsfxsize=16cm\epsfbox{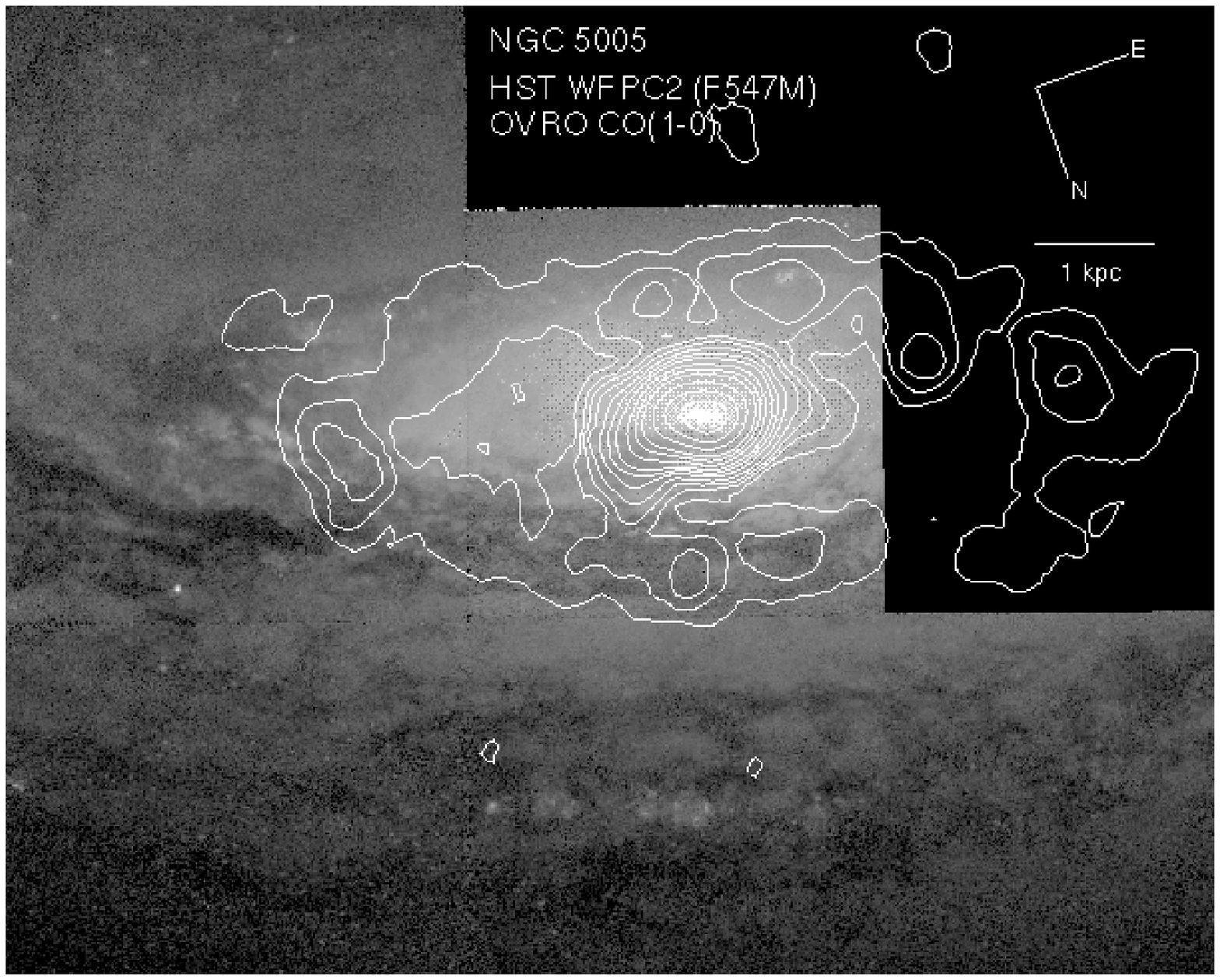}\hfill} 
\vspace{1cm}
\caption{Quality of this figure was lowered for astro-ph.
A PS preprint with the full-resolution figures is avaiable from
{\tt http://www.ovro.caltech.edu/mm/science/science.html} . }
\end{figure*}

\clearpage

\begin{figure*}[t]
{\hfill\epsfysize=8cm\epsfbox{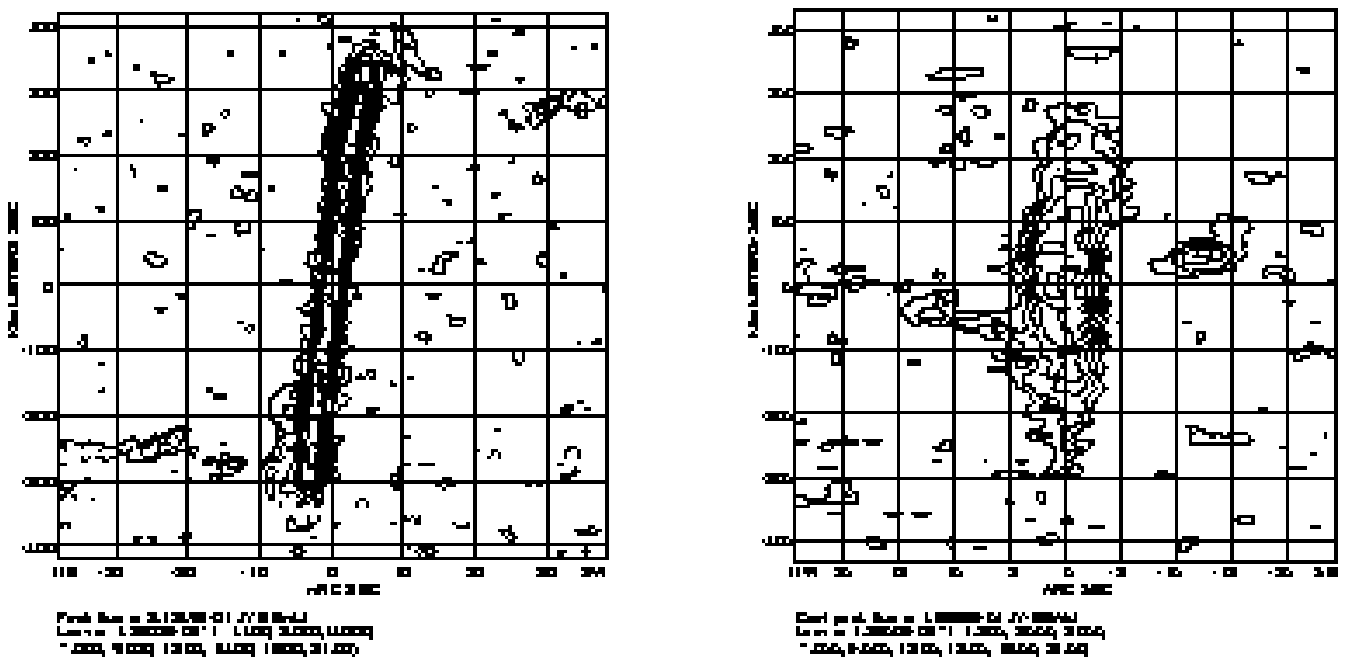}\hfill}
\vspace{1cm}
\caption{Quality of this figure was lowered for astro-ph.
A PS preprint with the full-resolution figures is avaiable from
{\tt http://www.ovro.caltech.edu/mm/science/science.html} . } 
\end{figure*}

\clearpage 

\begin{figure*}[t]
{\hfill\epsfxsize=15cm\epsfbox{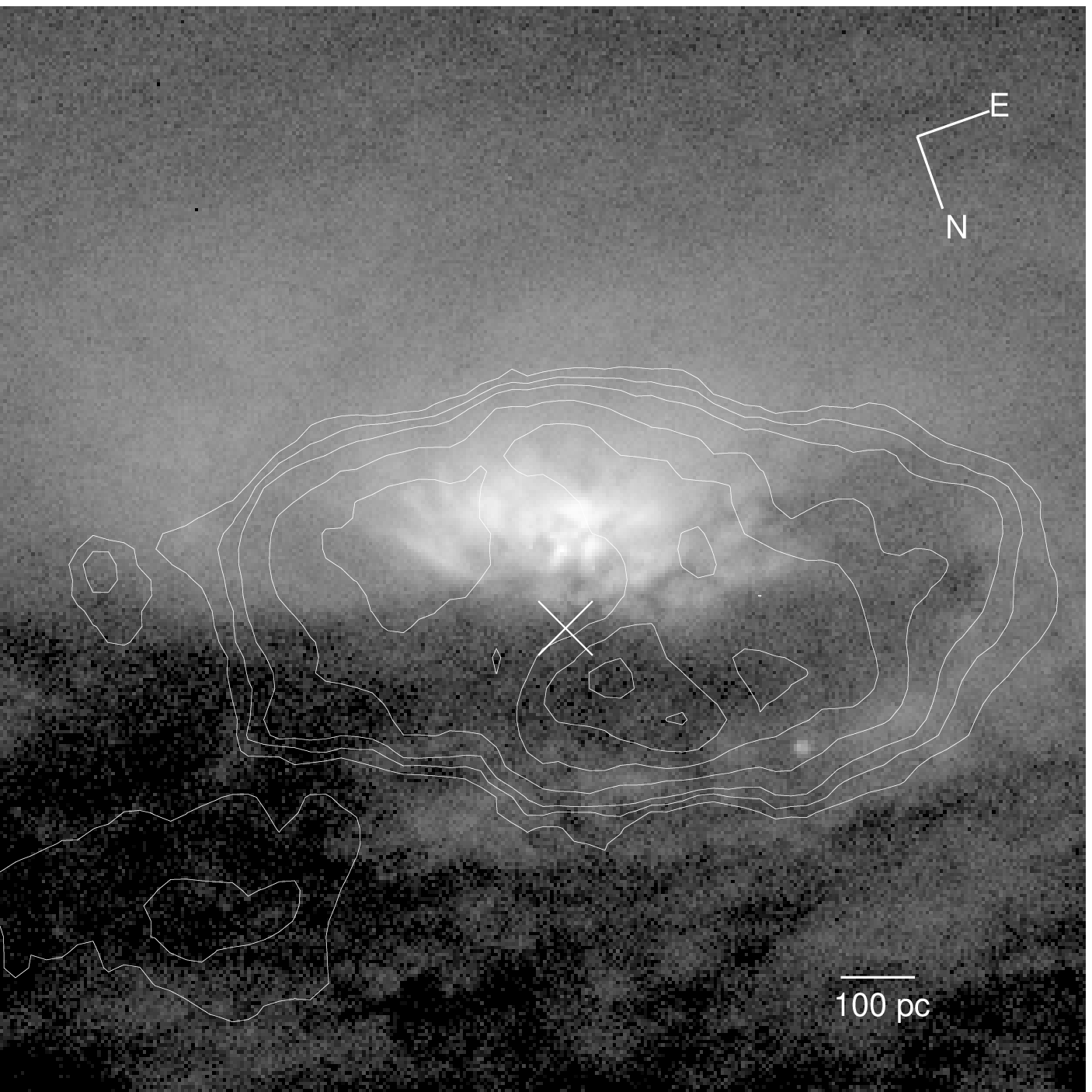}\hfill}
\vspace{1cm}
\caption{ } 
\end{figure*}

\clearpage

\begin{figure*}[t]
{\hfill\epsfxsize=17cm\epsfbox{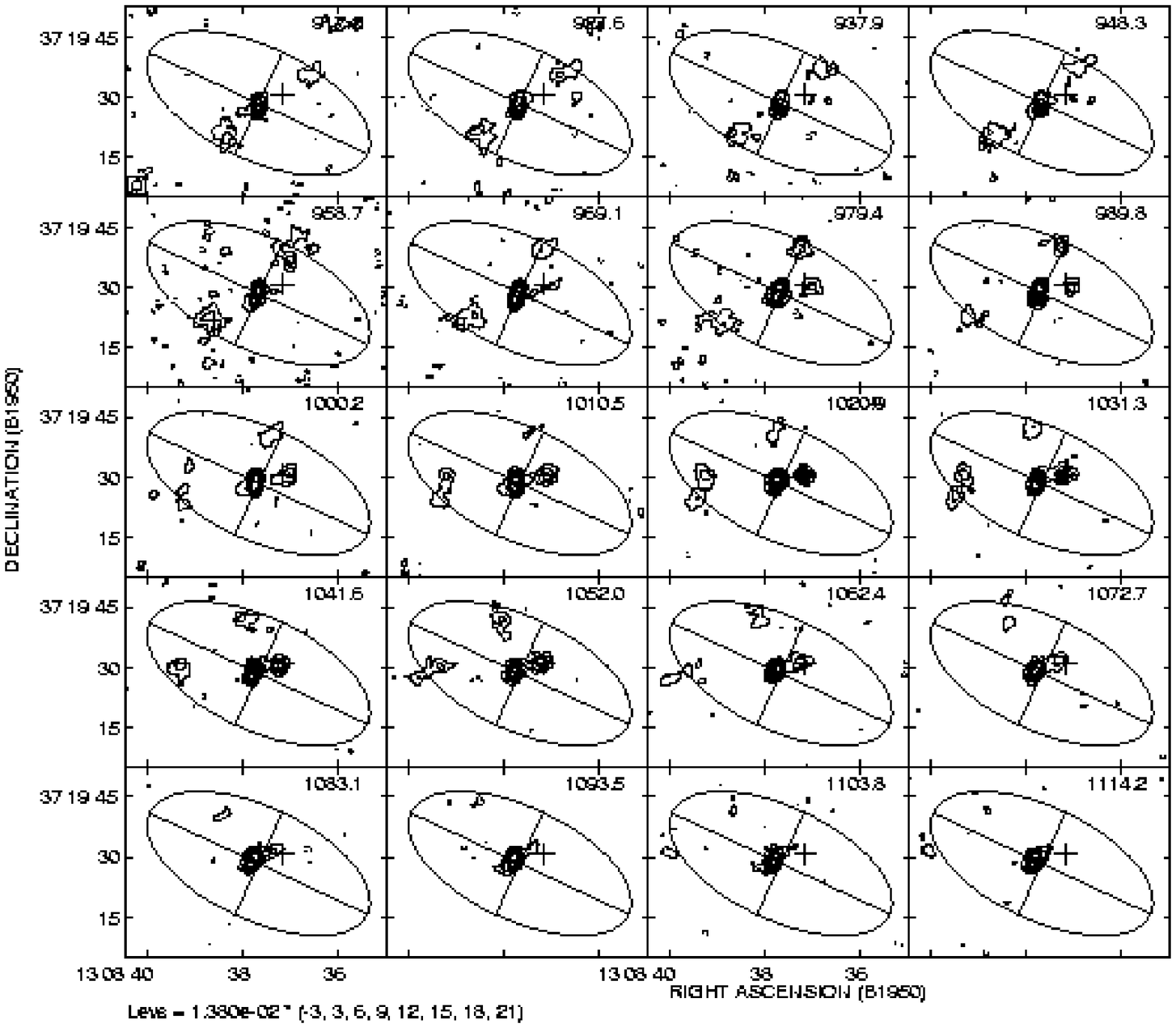}\hfill}
\vspace{1cm}
\caption{Quality of this figure was lowered for astro-ph.
A PS preprint with the full-resolution figures is avaiable from
{\tt http://www.ovro.caltech.edu/mm/science/science.html} .  }
\end{figure*}

\clearpage

\begin{figure*}[t]
{\hfill\epsfxsize=10cm\epsfbox{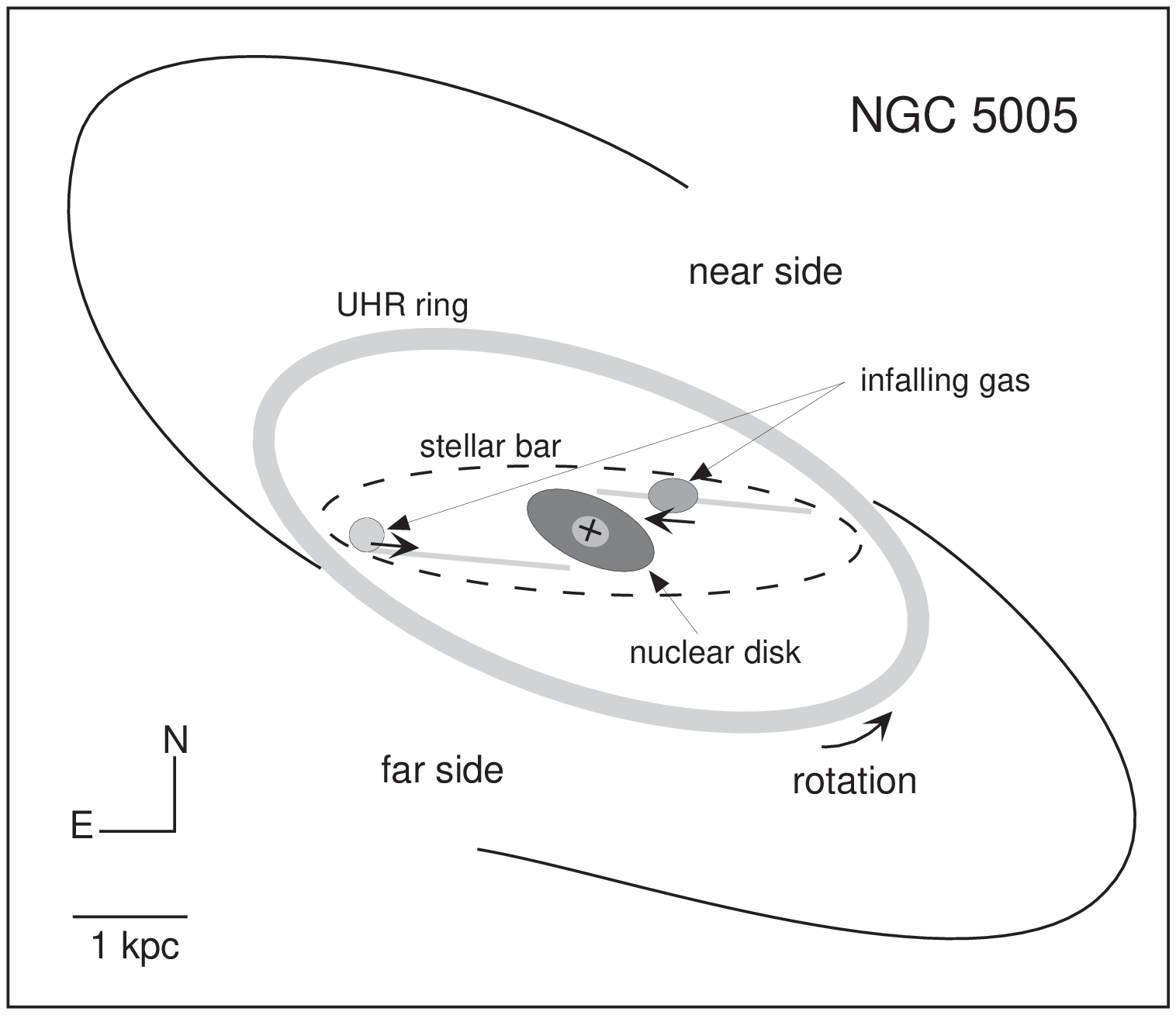}\hfill} 
\vspace{1cm}
\caption{ }
\end{figure*}

\clearpage

\begin{figure*}[t]
{\hfill\epsfxsize=7cm\epsfbox{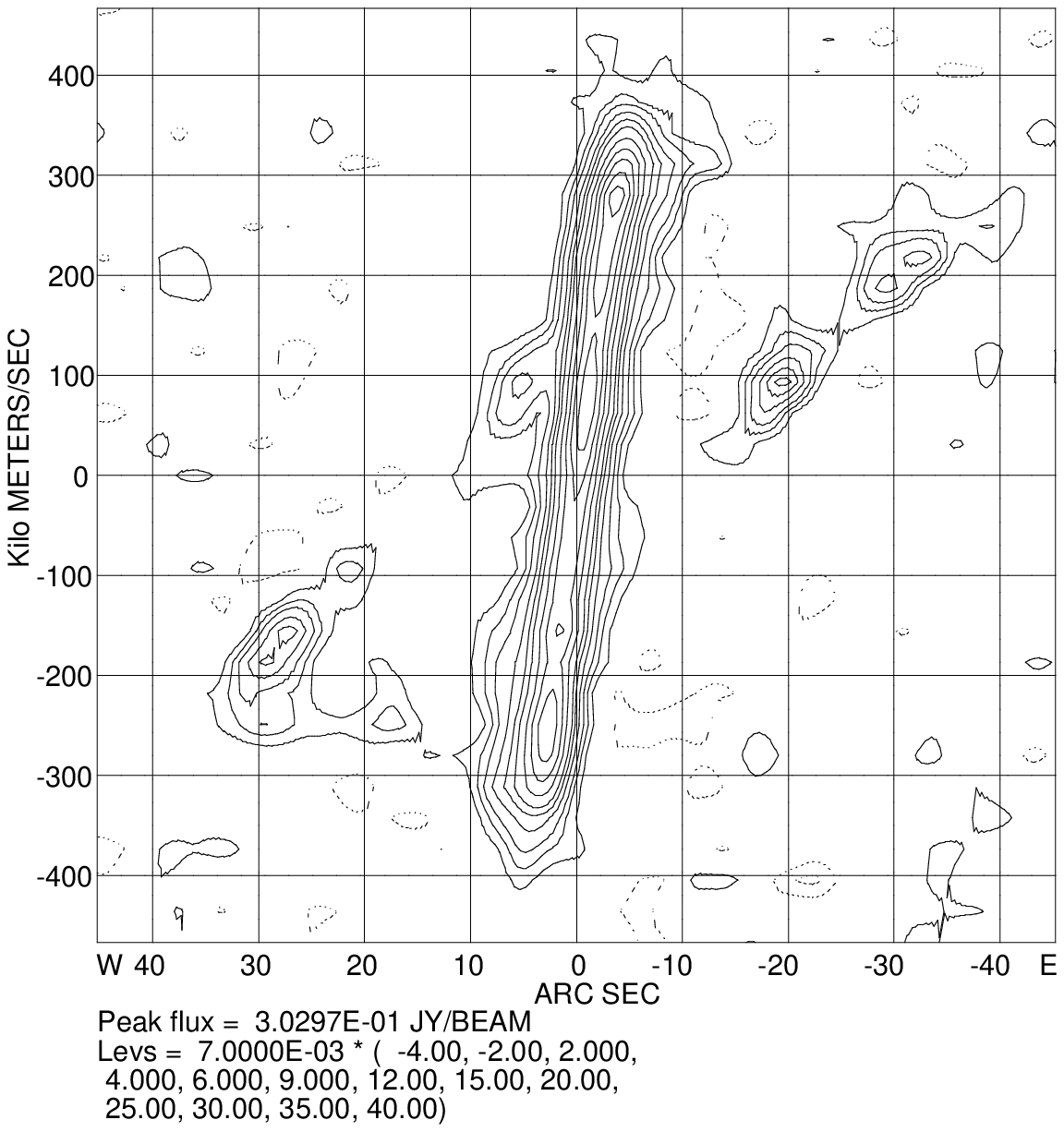}\hfill} 
\vspace{1cm}
\caption{ }
\end{figure*}


\begin{figure*}[t]
{\hfill
\epsfxsize=7cm\epsfbox{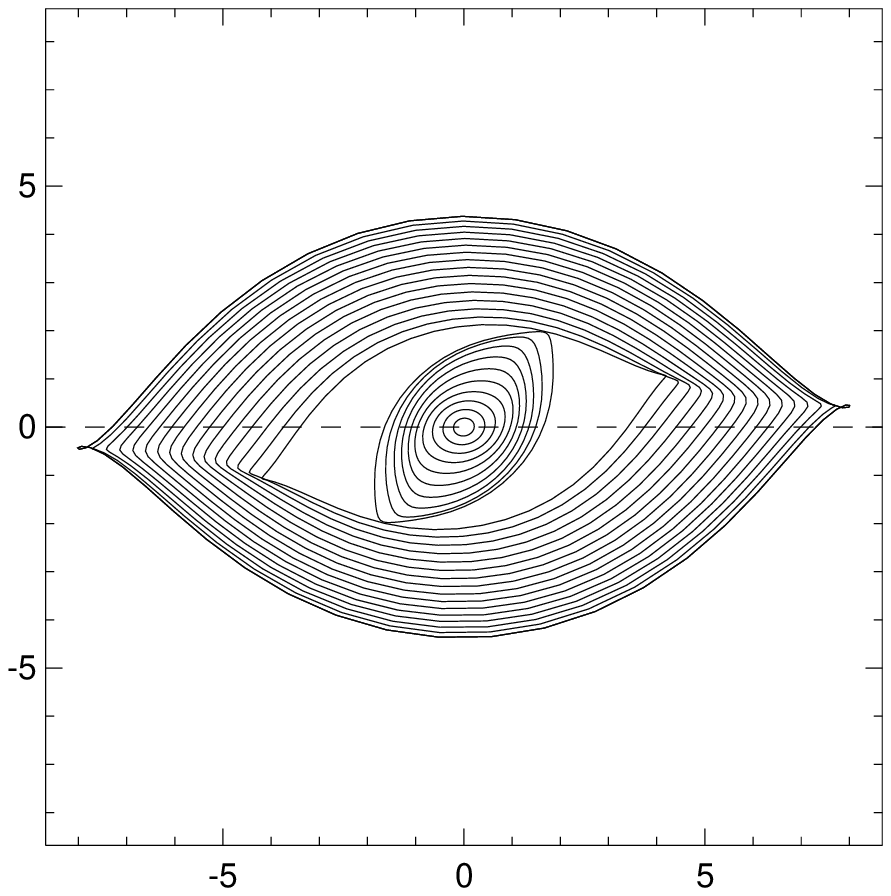}
\hspace{0.5cm}
\epsfxsize=7cm\epsfbox{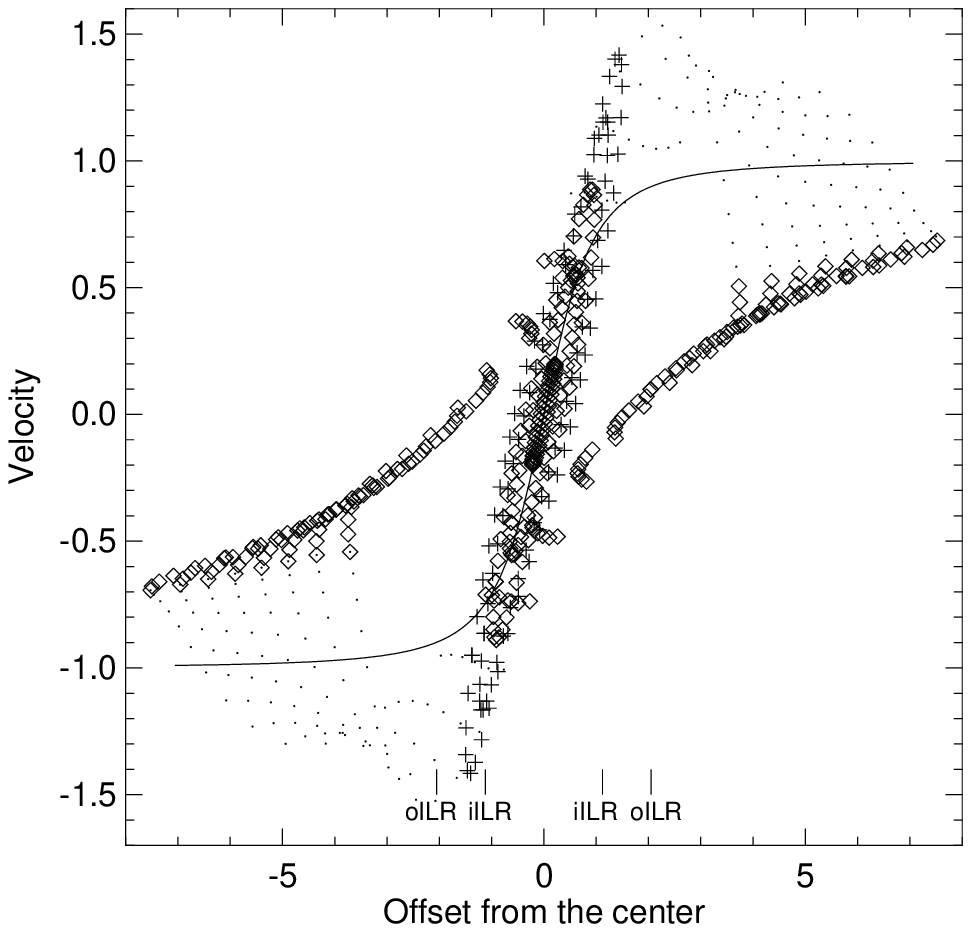}
\hfill}
\clearpage
\caption{ }
\end{figure*}

\clearpage

\begin{figure*}[t]
{\hfill\epsfxsize=8cm\epsfbox{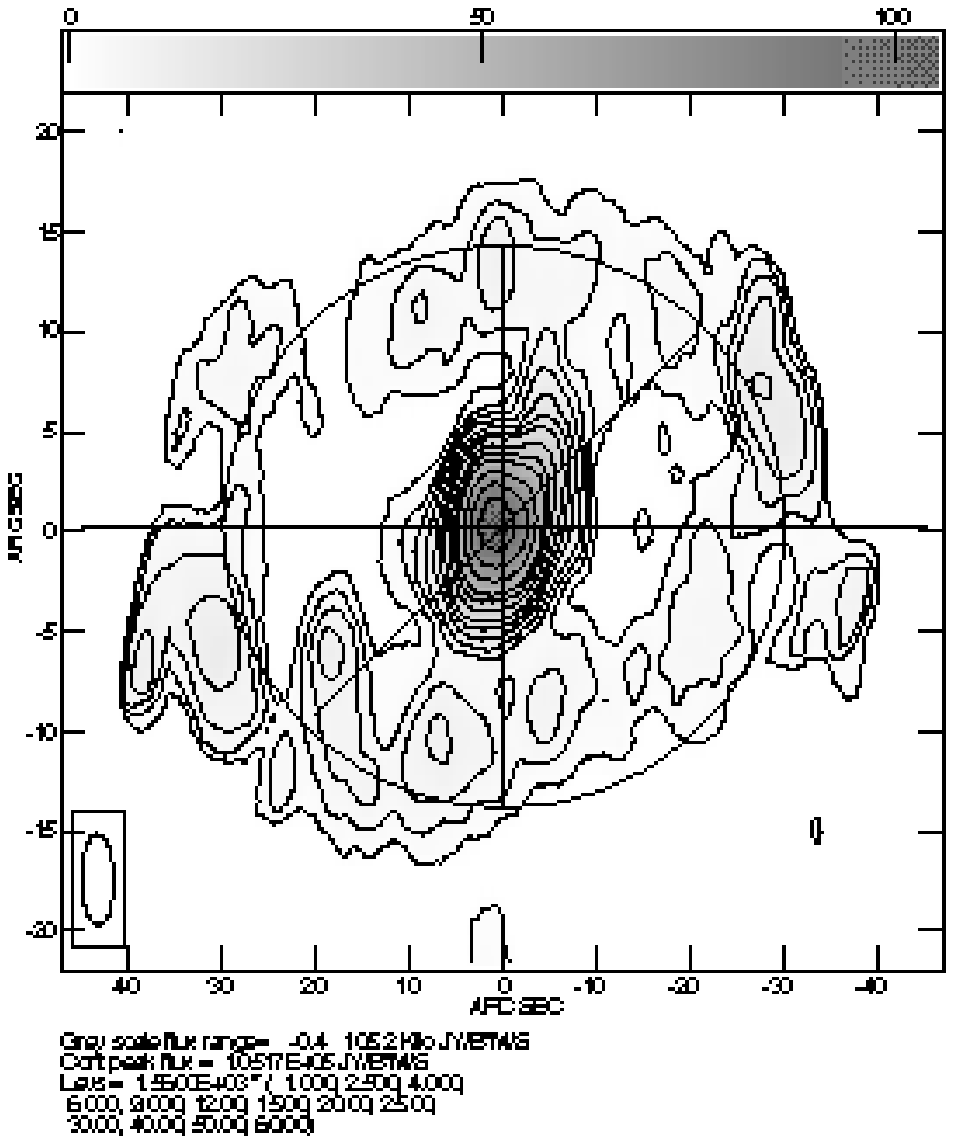}\hfill} 
\caption{Quality of this figure was lowered for astro-ph.
A PS preprint with the full-resolution figures is avaiable from
{\tt http://www.ovro.caltech.edu/mm/science/science.html} . }
\clearpage
\end{figure*}


\begin{figure*}[t]
{\hfill\epsfxsize=9cm\epsfbox{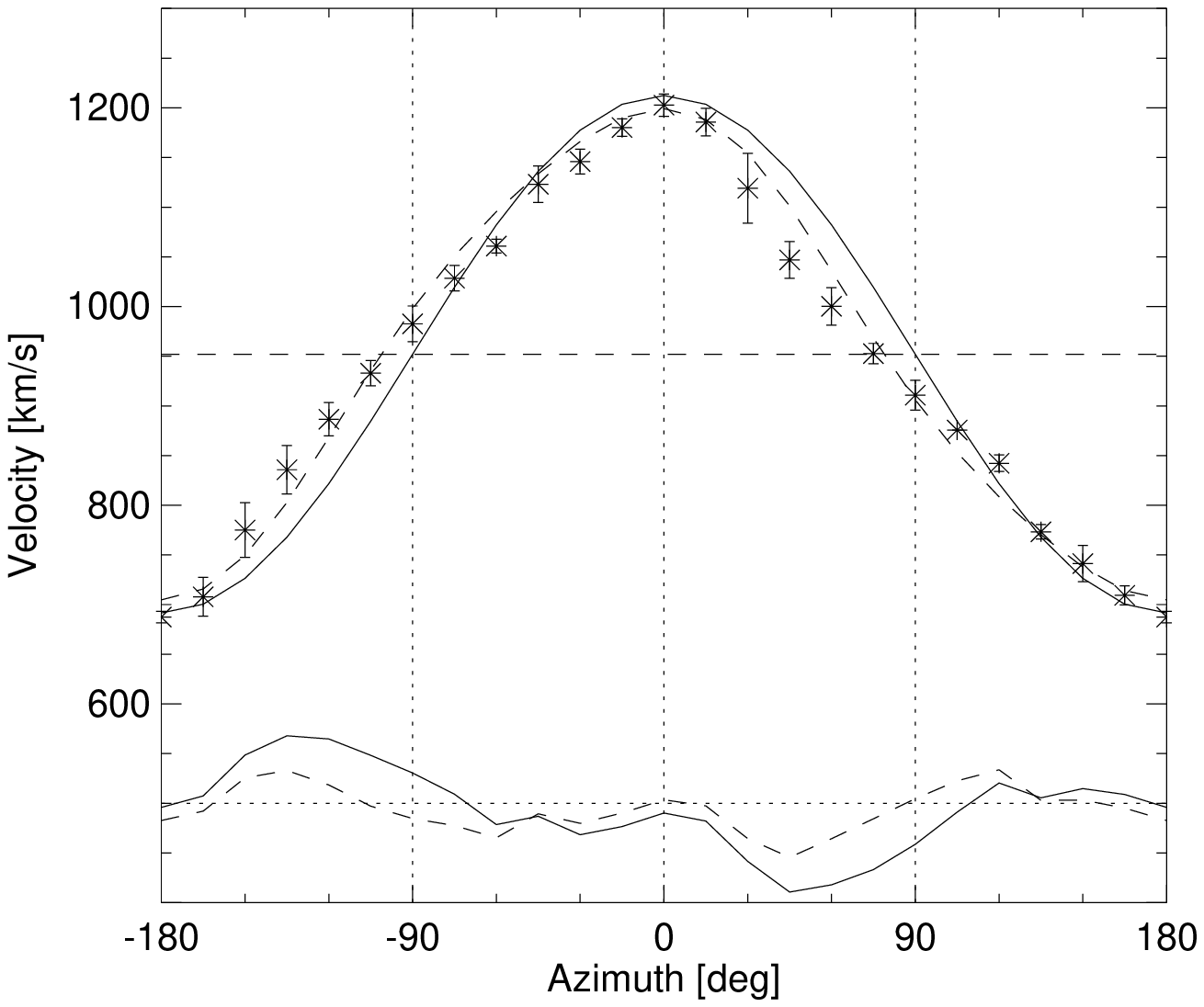}\hfill} 
\caption{ }
\clearpage
\end{figure*}

\end{document}